\documentclass[final,5p,times,twocolumn]{elsarticle}
\usepackage{amssymb}
\usepackage{amsmath}
\usepackage{amssymb}
\usepackage{psfrag}
\usepackage{color}
\usepackage{stfloats}
\usepackage{fixltx2e}
\usepackage{mathrsfs}
\usepackage{tabularx}
\usepackage{booktabs}
\usepackage{colortbl}
\usepackage{rotating}
\usepackage{boldline}
\usepackage{bm}
\usepackage{changes}
\usepackage{url}
\usepackage{adjustbox}
\usepackage{graphicx}
\usepackage{cellspace}
\usepackage{lineno}

\journal{arXiv}
\begin{document}
\renewcommand{\topfraction}{0.98}   
\renewcommand{\bottomfraction}{0.98}
\setcounter{topnumber}{3}
\setcounter{bottomnumber}{3}
\setcounter{totalnumber}{4}         
\setcounter{dbltopnumber}{4}        
\renewcommand{\dbltopfraction}{0.98}
\renewcommand{\textfraction}{0.05}  
\renewcommand{\floatpagefraction}{0.95}      
\renewcommand{\dblfloatpagefraction}{0.95}   
\newcommand{\beq}{\begin{equation}}
\newcommand{\eeq}{\end{equation}}
\newcommand{\D}  {\displaystyle}
\newcommand{\DS} {\displaystyle}
\def\sca   #1{\mbox{\rm{#1}}{}}
\def\mat   #1{\mbox{\bf #1}{}}
\def\vec   #1{\mbox{\boldmath $#1$}{}}
\def\scas  #1{\mbox{{\scriptsize{${\rm{#1}}$}}}{}}
\def\scaf  #1{\mbox{{\tiny{${\rm{#1}}$}}}{}}
\def\vecs  #1{\mbox{\boldmath{\scriptsize{$#1$}}}{}}
\def\tens  #1{\mbox{\boldmath{\scriptsize{$#1$}}}{}}
\def\tenf  #1{\mbox{{\sffamily{\bfseries {#1}}}}}
\def\ten   #1{\mbox{\boldmath $#1$}{}}
\sloppy
\begin{frontmatter}
\title{\Large{{\textsf{\textbf{Open-source benchmarking of plant-based and animal meats}}}}}
\author[inst1]{Sybren D. van den Bedem}
\author[inst2]{Ellen Kuhl\corref{cor1}}
\author[inst3]{Caroline Cotto \corref{cor1}}
\ead{syb@stanford.edu, 
     ekuhl@stanford.edu,
     caroline.cotto@nectar.org}
\cortext[cor1]{Corresponding authors}
\affiliation[inst1]{organization={Department of Mathematics, 
                                  Stanford University},
            city={Stanford, CA}, 
            country={United States}}
\affiliation[inst2]{organization={Department of Mechanical Engineering, 
                                  Stanford University},
            city={Stanford, CA}, 
            country={United States}}
\affiliation[inst3]{organization={NECTAR, 
                                  Food System Innovations},
            city={Claremont, CA}, 
            country={United States}}
\begin{abstract} %
Global food production must reduce environmental impact 
while meeting rising demand for dietary protein. 
Plant-based meats aim to preserve the sensory and cultural role of animal meat, while lowering greenhouse gas emissions, land use, and health risks. Advances in protein structure and flavor chemistry have improved product quality; yet, consumers continue to prioritize taste and texture over sustainability and systematic large-scale consumer surveys are scarce. It remains unclear how plant-based products rank against animal benchmarks and which product attributes most strongly influence overall liking.
Here we show, in a large-scale, blinded, in-person sensory evaluation 
across 14 product categories,
2,684 consumers,
more than 11,000 product evaluations and 800,000 data points, 
that plant-based products still trail animal benchmarks at the category average level, 
but approach parity in selected formats: 
Plant-based unbreaded chicken filets, 
chicken nuggets, and 
burgers achieved mean overall liking scores of 5.1, 4.9, and 5.2, differing from the animal benchmark 
by only $\Delta$ = 0.1, 0.2, and 0.3 points 
on a seven-point scale. 
For unbreaded chicken filets and burgers, 
48\% and 47\% of participants rated the plant-based product 
the same as or better than the animal benchmark. 
Categories with higher sensory parity captured 5\,$–$\,14\% market share 
compared with less than 1\% for low-parity categories. 
Penalty analysis identified savoriness, aftertaste, juiciness, and tenderness as the strongest determinants of liking.
These findings show that sensory parity is technically achievable, 
but not yet consistent across product types. 
By publicly sharing all sensory, preference, and market-linked data, 
we establish an open benchmark for alternative protein performance
to democratize research and accelerate principled, data-driven innovation.\\[6.pt]
All data are freely available at 
https://www.nectar.org/sensory-research/2025-taste-of-the-industry.
\end{abstract}
\begin{keyword}
sensory evaluation;
consumer acceptance;
plant-based meat;
alternative protein;
overall liking;
texture perception
\end{keyword}
\end{frontmatter}
\section{{\textsf{\textbf{Motivation}}}}
The global food system approaches a breaking point. Decades of research have already established that food production generates roughly one third of anthropogenic greenhouse gas emissions and accelerates deforestation, freshwater depletion, and biodiversity loss \cite{crippa21}. Global demand for food will rise sharply by mid-century, intensifying pressure on already stressed ecosystems \cite{tilman14}. Current dietary trajectories push planetary systems beyond safe operating limits and jeopardize climate stabilization goals \cite{poore18}. These facts no longer surprise; they define the baseline of contemporary food science. What remains unresolved is how to translate this knowledge into scalable dietary change. \\[6.pt]
{{\textsf{\textbf{Animal agriculture amplifies environmental and health risk.}}}}
Livestock production contributes substantially to greenhouse gas emissions, land degradation, and water scarcity \cite{clark20}. Beef production alone carries a carbon footprint many times higher than most plant-based protein sources \cite{poore18}. Diets high in red and processed meat associate with elevated risks of cardiovascular disease, type 2 diabetes, and colorectal cancer \cite{willett19}. Modeling studies show that widespread dietary shifts toward plant-forward patterns could reduce food-related greenhouse gas emissions by up to 70\% while improving global health outcomes \cite{springmann18}. Policymakers increasingly recognize dietary transition as essential for meeting Paris Agreement targets \cite{ipcc22}. Yet, societies will not abandon meat unless alternatives satisfy deeply rooted expectations for taste, texture, and culinary experience \cite{godfray18}.\\[6.pt]
{{\textsf{\textbf{Plant-based meats redefine meat beyond the animal.}}}}
Plant-based products seek to preserve the sensory and cultural role of meat while eliminating the environmental burden of animal agriculture \cite{joshi15}. Modern formulations combine structured plant proteins, fats, binders, and flavor systems to approximate muscle architecture and mouthfeel \cite{kyriakopoulou19}. High-moisture extrusion and shear-cell technologies now generate fibrous textures that increasingly resemble animal muscle tissue \cite{dekkers18}. Life-cycle assessments demonstrate substantial reductions in greenhouse gas emissions, land use, and water use relative to conventional beef \cite{heller18}. However, sensory performance varies markedly across products and categories \cite{elzerman11}. Although many consumers express concern about environmental sustainability, flavor and texture remain the primary determinants of food choice and drive preference for animal meat over plant-based alternatives \cite{hoek11}. Researchers seek to address this sensory gap by linking measurable mechanical properties \cite{dunne25} to perceived texture \cite{szczesniak02}, but product development still relies largely on iterative reformulation rather than systematic, transparent benchmarking against animal counterparts \cite{kuhl25}. \\[6.pt]
{{\textsf{\textbf{Consumers remain skeptical.}}}}
Adoption depends on sensory satisfaction, not environmental intent \cite{bryant18}. Many consumers perceive plant-based meats as inferior in savoriness, juiciness, and overall flavor depth \cite{stpierre24}. Sensory analyses reveal that meat is consistently associated with positive terms whereas meat alternatives are viewed more negatively \cite{michel21}. These findings clearly define the performance gap \cite{sogari23}, and major food manufacturers invest heavily in reformulation efforts to close it. Yet, most large-scale sensory datasets remain proprietary and siloed within companies, which limits independent validation, cross-category comparison, and cumulative scientific progress \cite{mcclements21}. This fragmentation obscures true performance, blunts accountability, and slows the path toward sensory parity \cite{datta25}.\\[6.pt]
Here we address these limitations by conducting a large-scale, blinded sensory comparison of plant-based and animal meats across multiple categories to establish a rigorous and transparent benchmark of current performance \cite{nectar24}. We share all sensory data and analyses as an open source resource to democratize food science and accelerate discovery and innovation toward sustainable protein systems \cite{nectar25}.
\section{{\textsf{\textbf{Methods}}}}
We conducted a large-scale, blinded, in-person sensory study 
to compare plant-based meat products 
with conventional animal benchmarks 
across 14 product categories
between November 2024 and January 2025 \cite{nectar25}.
To reflect real-world consumption contexts
we performed all tests at restaurant partner locations 
in San Francisco, CA and New York City, NY.
We prepared all products according to manufacturer instructions and
allowed participants to add condiments when appropriate.
We implemented a blinded, randomized, within-subject design, where
participants evaluated one product at a time. 
After tasting each product, 
participants completed a standardized survey 
that captured sensory evaluations, 
similarity to conventional products, 
purchase intent, and open-ended feedback questions. 
The full study included 
2,684 participants,
more than 11,000 plant-based product evaluations,
and more than 800,000 data points. 
\subsection{{\textsf{\textbf{Tested products}}}}
\begin{figure*}[h]
\centering
\includegraphics[width=0.75\linewidth]{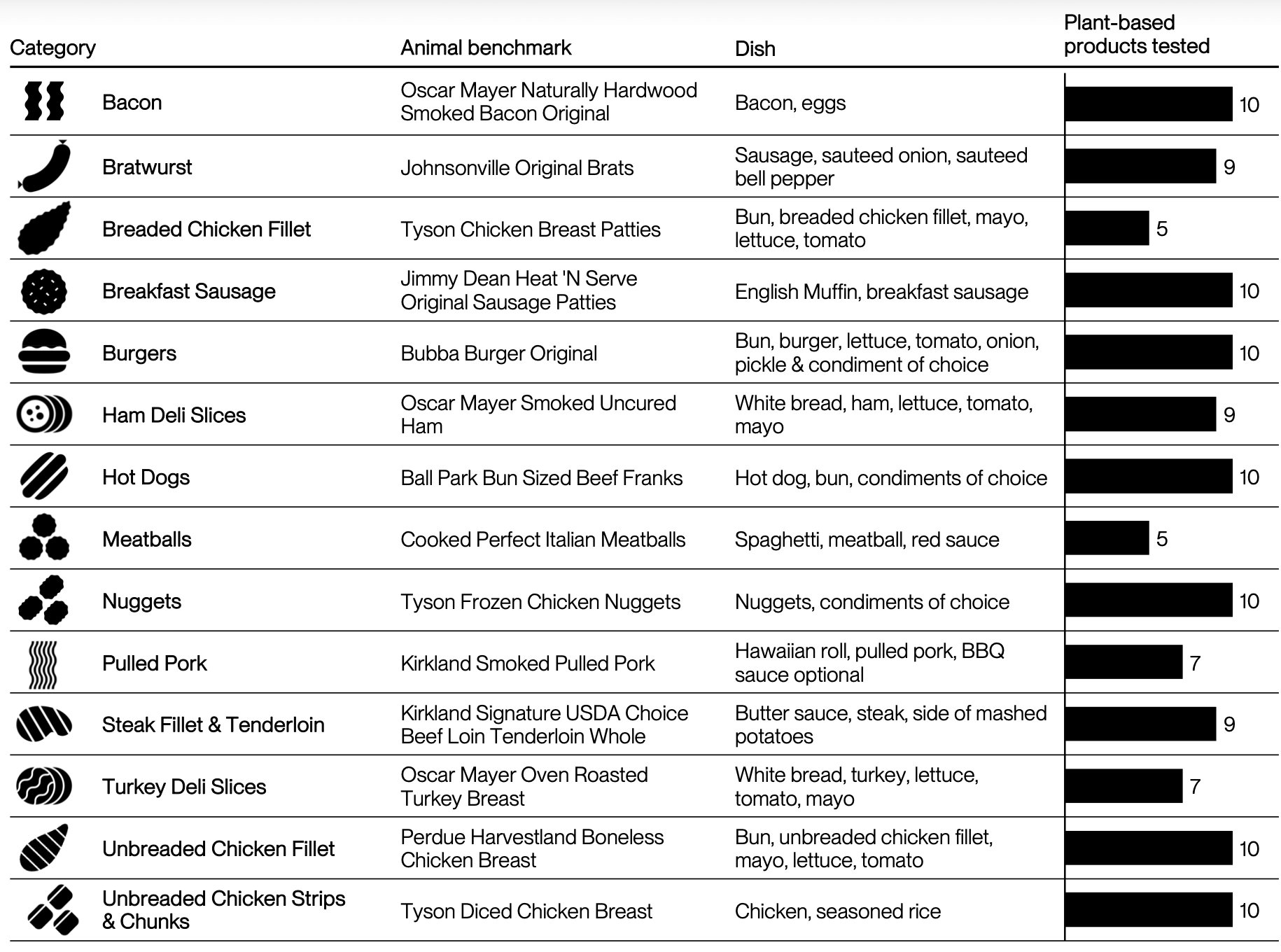}
\caption{{\bf{\sffamily{Tested products.}}} This study tested 14 product categories with a total 121 plant-based products and 14 animal benchmarks. 
We selected the 14 product categories by high sales volume and market development, and the 121 plant-based products by market presence, market-readiness, animal-analog design, animal-compatible flavor, and distinct ingredient or production technology. We served each product in a complete but simplified product-specific dish.}
\label{fig01}
\end{figure*}
We selected fourteen product categories based on two criteria: 
high sales volume and 
sufficient market development with at least five plant-based products available within the category (Fig. \ref{fig01}). 
Within these categories, 
we prioritized 121 plant-based products 
using the following criteria:
significant market presence,
market-ready status, 
design as animal analog,
original flavor comparable to animal product,
distinct ingredients or production technology 
relative to other products tested in the category.
For each category, 
we selected a conventional animal benchmark 
based on highest retail sales volume. 
The fourteen categories include 
bacon, 
bratwurst, 
breaded chicken filet,
breakfast sausage, 
burger, 
chicken nuggets,
deli ham, 
deli turkey,
hot dogs, 
meatballs, 
pulled pork, 
steak, 
unbreaded chicken filet, and
unbreaded chicken strips. 
We served each item as a complete but simplified build 
that matched typical consumption formats for the category.
\subsection{{\textsf{\textbf{Study population}}}}
\begin{figure}[h]
\centering
\includegraphics[width=1.0\linewidth]{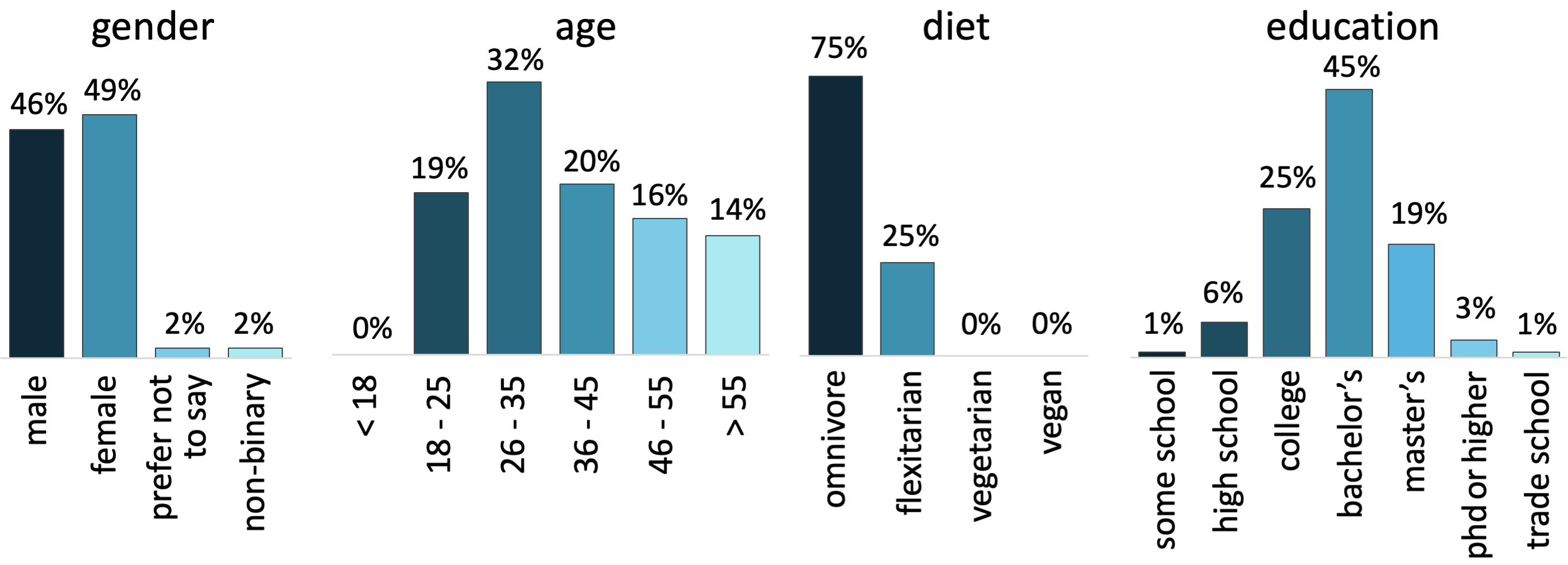}
\caption{{\bf{\sffamily{Study population.}}} 
Demographic overview of the 2,684 study participants.
We enrolled a gender-, age-, and educationally balanced population 
of omnivores and flexitarians
to ensure that all participants actively consumed animal meat.}
\label{fig02}
\end{figure}
The study population consists 
of 2,684 U.S. consumers,
of whom 75\% identified as omnivores and 25\% as flexitarians (Fig. \ref{fig02}).
The gender distribution includes
49\% female, 46\% male, 2\% non-binary, and 2\% who preferred not to disclose.
The age distribution includes
19\% 18–25 years old, 
32\% 26–35,
20\% 36–45,
16\% 46–55, and 
14\% over 55.
The study population had a diverse education background with 
 6\% holding a high school degree,
25\% a college degree,
45\% a bachelor's degree,
19\% a master's degree, 
 3\% a Ph.D. or higher, and
 3\% some other school. 
We excluded vegetarians and vegans 
to ensure that all participants actively consumed animal meat 
and can provide meaningful benchmark comparisons.
\subsection{{\textsf{\textbf{Data analysis}}}}
Participants rated overall liking, flavor, texture, and appearance on a seven-point Likert scale ranging from dislike very much (1) to like very much (7) and rated purchase intent from definitely would not buy (1) to definitely would buy (7).
We calculated mean scores for each product 
and compared three categories, 
the {\it{plant-based leader}},
defined as the highest-rated plant-based product per category,
the {\it{plant-based average}},
defined as the average per category, and
the {\it{animal benchmark}}
using Wilcoxon signed-rank tests and 
report statistical significance based on two-sided p-values.
We group the results of the overall liking ratings
into {\it{promoters}}, including like very much (7) and like (6),
{\it{passive}}, including like somewhat (5) and neither like nor dislike (4), and
{\it{detractors}}, including dislike somewhat (3), dislike (2), and dislike very much (1).
Participants also completed check-all-that-apply questions that covered 
flavor, texture, and appearance attributes
from which we calculated 
attribute prevalence and conducted penalty analysis 
with mean drop and lift 
to quantify associations between attributes and overall liking. 
\section{{\textsf{\textbf{Results}}}}
\begin{figure*}[h]
\centering
\includegraphics[width=0.75\linewidth]{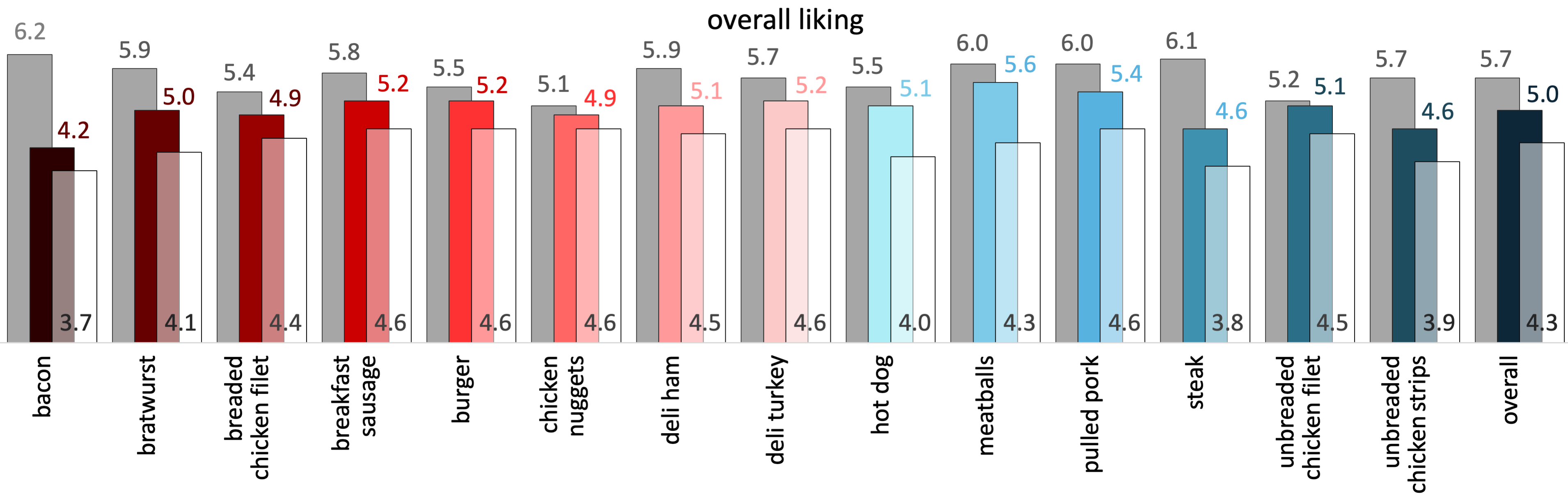}
\caption{{\bf{\sffamily{Overall liking scores.}}}
Mean overall liking scores of
animal benchmark (gray),
plant-based leader (dark color), and
plant-based average (white) 
across all 14 product categories.
Approximately 100 participants evaluated each product 
(mean per test = 99.4) on a seven-point Likert scale
from dislike very much (1) to like very much (7). 
The overall mean across all 14 categories 
ranged from
5.7 for animal benchmark, to 
5.0 for plant-based leader, and
4.3 for plant-based average (right).}
\label{fig03}
\end{figure*}
In all 14 product categories, the animal benchmark received the highest overall liking scores, the plant-based leader received the second highest overall liking scores, and the plant-based average ranked lowest (Fig. \ref{fig03}). The taste gap between the animal benchmark and the leading plant-based product was smallest in unbreaded chicken filet ($\Delta = 0.1$) and chicken nuggets ($\Delta = 0.2).$ The taste gap was largest for bacon ($\Delta = 2.0$) and steak ($\Delta = 1.5$). The mean overall liking was 5.7 for the animal benchmark, 5.0 for the plant-based leader, and 4.3 for the plant-based average, which shows a clear gradient in consumer acceptance. These results indicate that leading plant-based products can approach animal benchmarks in overall liking in selected categories, but they do not yet match animal products on overall liking at the category level.
\begin{figure*}[h]
\centering
\includegraphics[width=0.75\linewidth]{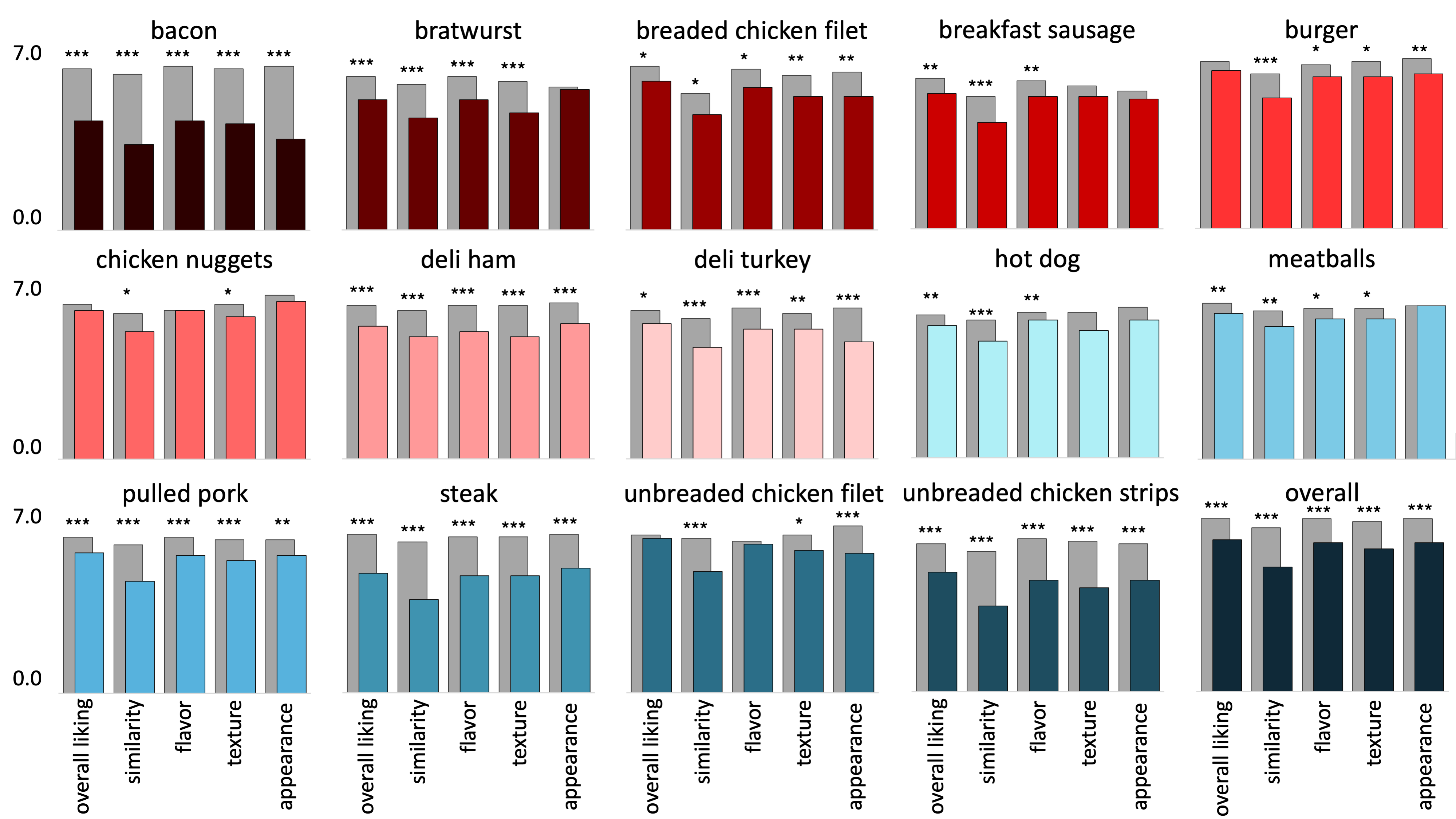}
\caption{{\bf{\sffamily{Overall liking, similarity, flavor, texture, and appearance scores.}}}
Performance of
animal benchmark (gray) and
plant-based leader (dark color)
across all 14 product categories,
bacon, bratwurst, breaded chicken filet, breakfast sausage, burger, 
chicken nuggets, deli ham, deli turkey, hot dog, meatballs,
pulled pork, steak, unbreaded chicken filet, and unbreaded chicken strips,
scored by
mean overall liking, similarity, flavor, texture, and appearance.
Approximately 100 participants evaluated each product 
(mean per test = 99.4) on a seven-point Likert scale
from dislike very much (1) to like very much (7). 
Across all 5 $\times$ 14 rankings,
in 11 categories, 
there was no significant difference between animal benchmark and plant leader;
in 59 categories, 
the animal benchmark scored significantly higher than the plant-based leader
$^{***}$p $>$ 0.001, $^{**}$p $>$ 0.01, $^{*}$ p$>$ 0.05.}
\label{fig04}
\end{figure*} \\[6.pt]
Across 70 category-by-attribute comparisons, the animal benchmarks scored significantly higher than the plant-based leaders in 59 cases, while 11 cases showed no significant comparisons (Fig. \ref{fig04}). No comparisons favored the plant-based products. This pattern was observed across all attributes, suggesting that the performance gap remains broad, rather than limited to single attributes. Similarity showed some of the largest and most consistent gaps across categories, which suggests that the plant-based products struggle to closely replicate the animal benchmarks. The eleven cases that showed no significant differences were spread across four attributes (overall liking, flavor, texture, appearance) and seven categories (bratwurst, breakfast sausage, burgers, pulled pork, hot dog, meatballs and unbreaded chicken filet). For attributes that showed a significant difference, the magnitude of the gap varied by category and attribute, suggesting that certain animal products may be more successfully replicated than others. These results show that leading plant-based products can approach animal benchmarks in specific category-attribute comparisons, but overall, the animal benchmark maintains an advantage across all categories.
\begin{figure*}[h]
\centering
\includegraphics[width=0.75\linewidth]{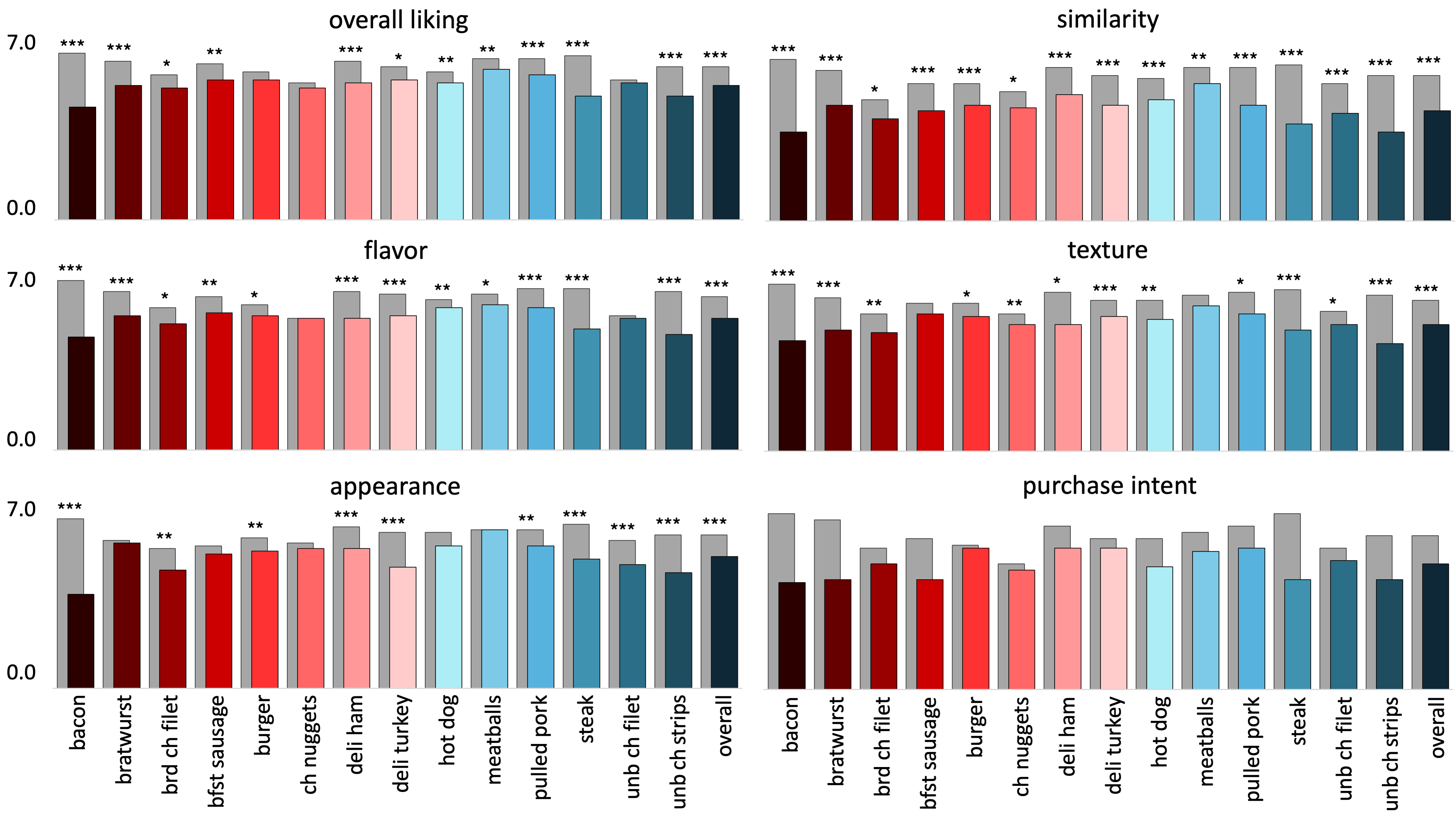}
\caption{{\bf{\sffamily{Overall liking, similarity, flavor, texture, appearance, and purchase intent scores.}}}
Performance of
animal benchmark (gray) and
plant-based leader (dark color)
scored by
mean overall liking, similarity, flavor, texture, appearance, and purchase intent
across all 14 product categories,
bacon, bratwurst, breaded chicken filet, breakfast sausage, burger, 
chicken nuggets, deli ham, deli turkey, hot dog, meatballs,
pulled pork, steak, unbreaded chicken filet, and unbreaded chicken strips.
Approximately 100 participants evaluated each product 
(mean per test = 99.4) on a seven-point Likert scale
from dislike very much (1) to like very much (7). 
Across all 5 $\times$ 14 rankings,
in 11 categories, 
there was no significant difference between animal benchmark and plant leader;
in 59 categories, 
the animal benchmark scored significantly higher than the plant-based leader
$^{***}$p $>$ 0.001, $^{**}$p $>$ 0.01, $^{*}$ p$>$ 0.05.}
\label{fig05}
\end{figure*} \\[6.pt]
Consistent with the sensory and liking difference (Fig. 4), purchase intent for the animal benchmark was numerically greater than the leading plant-based product across all 14 categories (Fig. \ref{fig05}). Thus, the performance gap was not limited to product evaluation, but extended to stated consumer willingness to purchase. The animal-to-plant-based difference varied across categories,
from bacon ($\Delta = 2.2$) and steak ($\Delta = 2.1$)
with the largest differences 
to chicken nuggets ($\Delta = 0.2$) and burgers ($\Delta = 0.1$)
with the smallest differences. 
This suggests 
that some product categories 
offer stronger commercial potential than others, 
although none reached parity with the animal benchmark. 
\begin{figure}[h]
\centering
\includegraphics[width=0.8\linewidth]{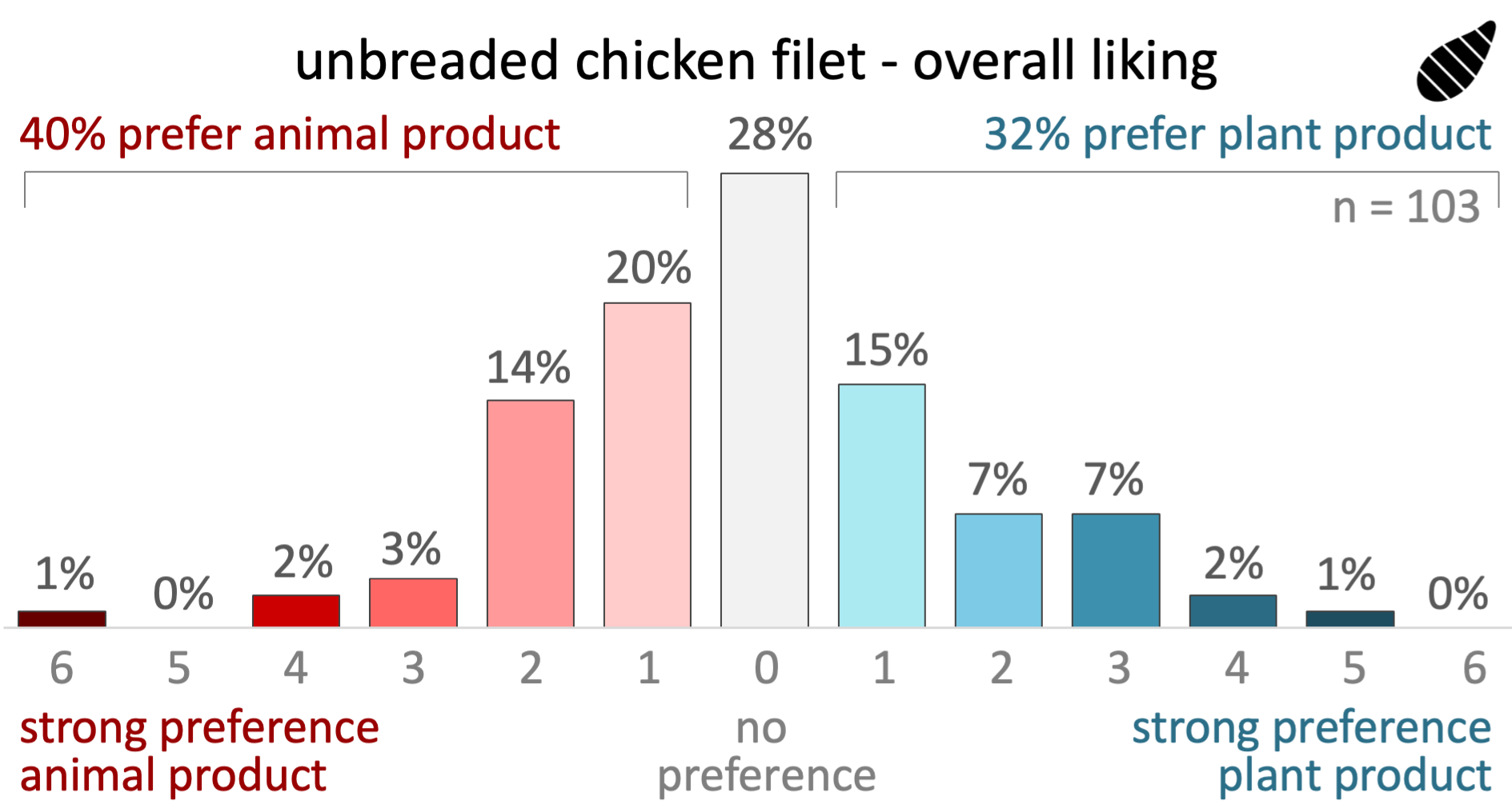}
\caption{{\bf{\sffamily{Distribution of paired differences in overall liking between plant-based leader and animal benchmark.}}}
Histogram of within-subject differences in seven-point overall liking score for Impossible Unbreaded Chicken Filets versus  
Perdue animal benchmark (n = 103). 
Positive values indicate higher ratings for the plant-based product; negative values indicate higher ratings for the animal product; zero indicates no difference; $40\%$ of participants preferred the animal benchmark, 32\% preferred the plant-based product, and 28\% reported no difference. The Wilcoxon signed-rank test detected no statistically significant preference for the animal product, p = 0.314.}
\label{fig06}
\end{figure} \\[6.pt]
The best performing plant-based category 
was unbreaded chicken
with the Impossible Unbreaded Chicken Filets
as plant-based leader (\ref{fig06}).
We calculated within-subject differences in overall liking 
by subtracting each participant's rating 
of the animal product 
from their rating of the plant-based product (n = 103). 
40\% of respondents rated the animal benchmark higher, 
32\% rated the plant-based product higher, and 
28\% reported no preference. 
The histogram also shows 
that many participants clustered near small differences, 
which indicates that many respondents 
considered the products as close in overall liking. 
Although the animal product had a descriptive advantage, 
preferences were broadly split, 
with over one-quarter of respondents indicating no preference. 
The Wilcoxon signed-rank test detected 
no significant overall preference for the animal benchmark ($p = 0.314$). 
These results indicate that that, 
for this product category, 
the leading plant-based product approached the animal benchmark 
which indicates that taste parity for this category lies within reach.
\begin{figure*}[h]
\centering
\includegraphics[width=0.8\linewidth]{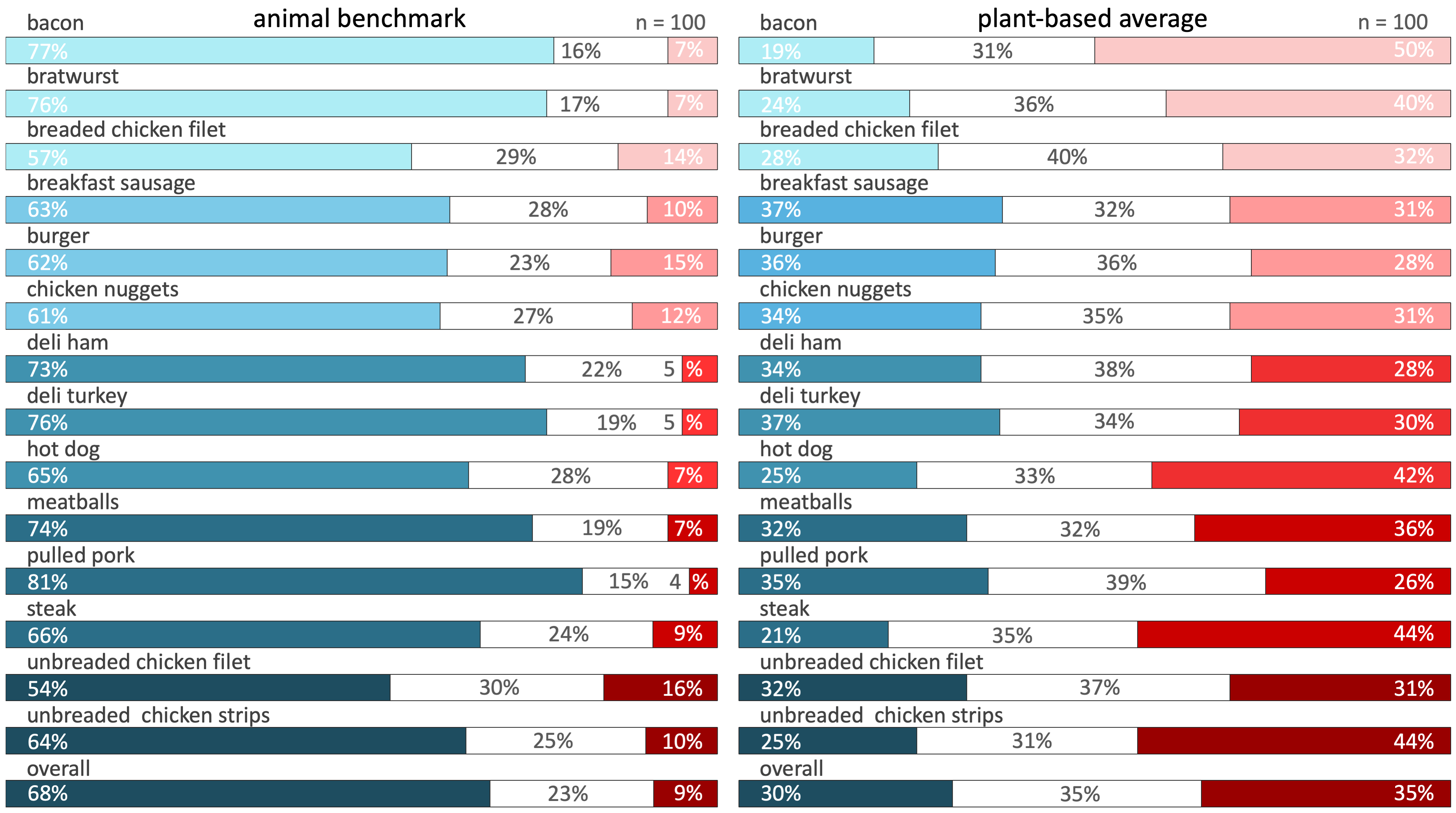}
\caption{{\bf{\sffamily{Promoter, passive, and detractor shares for plant-based averages and animal benchmarks across categories.}}} 
Stacked bars show the percentage of responses 
classified as promoters (like very much and like), 
passives (like somewhat and neither like or dislike), and 
detractors (dislike somewhat, dislike, dislike very much)
for each product category (n = 100 per test).
Across all categories, 
plant-based averages (right) show lower promoter shares 
and higher detractor shares than animal benchmarks (left). 
Overall, 30\% rated the plant-based average as promoters 
compared with 68\% for animal benchmarks, 
while detractor shares reached 35\% for plant-based products 
compared with 9\% for animal products.}
\label{fig07}
\end{figure*} \\[6.pt]
The overall liking ratings 
shows clear differences 
between plant-based averages 
and animal benchmarks across all categories (Fig. \ref{fig07}). 
We classified responses as
promoters (like very much and like), 
passives (like somewhat and neither like or dislike), and 
detractors (dislike somewhat, dislike, dislike very much)
based on seven-point overall liking ratings. 
Across all categories, 
plant-based products generated 
substantially fewer promoters and 
more detractors than animal benchmarks. 
On average, 
30\% of respondents rated the plant-based product as 
like very much or like, whereas 
68\% rated the animal benchmark in these top two categories. 
In contrast, 
35\% rated plant-based products in one of the dislike categories compared with 
9\% for animal products. 
The data also reveal wide variation 
across plant-based categories. 
Categories such as 
pulled pork, 
deli turkey, 
burgers, and 
breakfast sausages achieved promoter shares in the mid-30\% range, 
whereas 
bacon and steak filet showed 
promoter shares near or below 20\% and 
detractor shares exceeding 40\%. 
Taken together, 
these patterns highlight both 
the magnitude of the liking gap relative to animal benchmark 
and the uneven performance of plant-based products across categories.
\begin{figure}[h]
\centering
\includegraphics[width=0.8\linewidth]{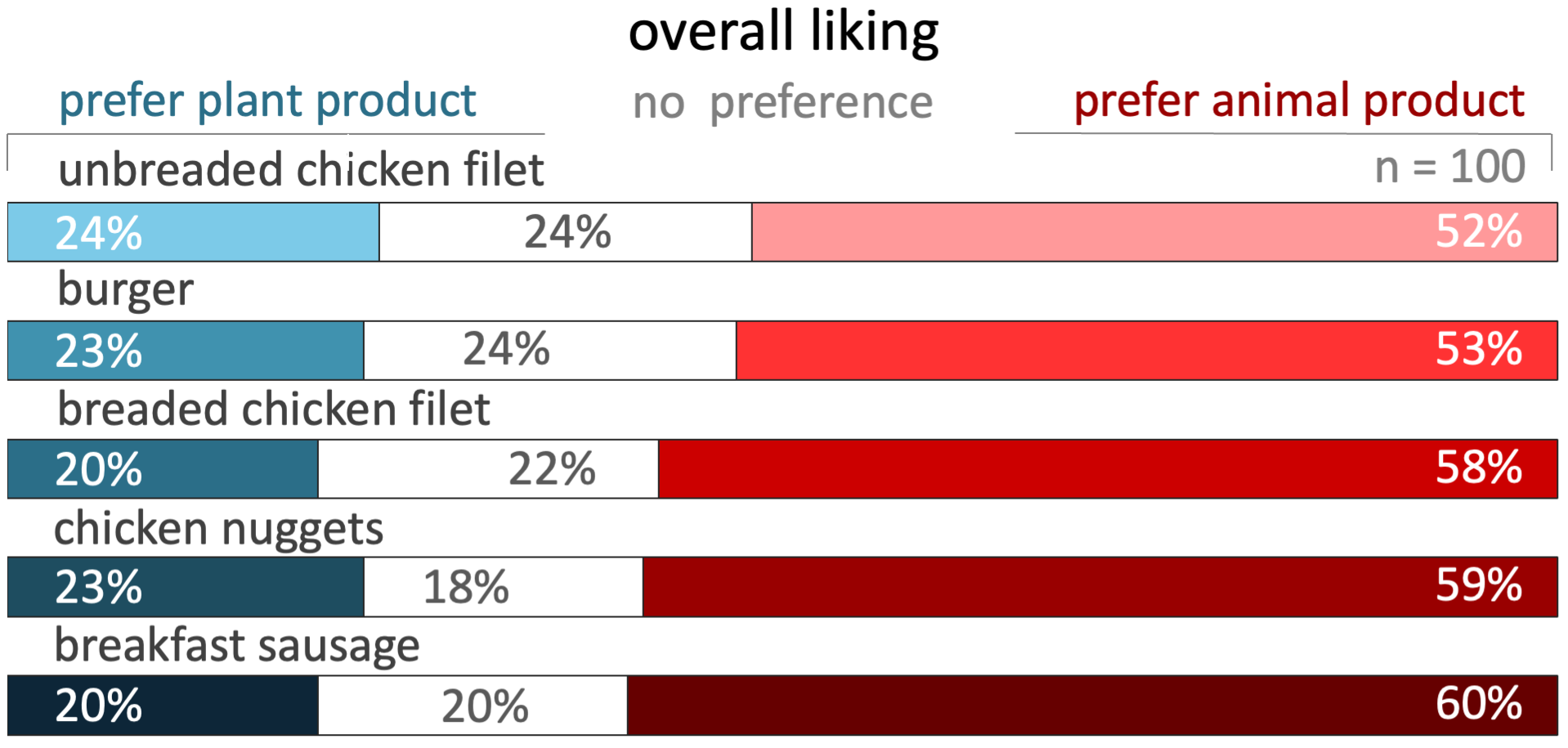}
\caption{{\bf{\sffamily{Preference comparison between plant-based leader and animal benchmark in top-performing categories.}}} 
Stacked bars show the percentage of participants who preferred the plant-based leader, reported no preference, or preferred the animal benchmark based on within-subject overall liking comparisons (n = 100 per test). 
Across the five highest-performing categories, 
at least 40\% of participants rated the plant-based product the same as or better than the animal benchmark. 
Unbreaded chicken filet (48\%) and burgers (47\%) achieved the highest shares of same or better ratings, 
followed by breaded chicken filet (42\%), nuggets (41\%), and breakfast sausages (40\%).}
\label{fig08}
\end{figure} \\[6.pt]
Five plant-based categories performed well against the animal benchmark
(Fig. \ref{fig08}).
For each category, 
we calculated preference 
by comparing each participant's overall liking ratings 
for the plant-based product against its animal benchmark. 
Across these five categories, 
at least 40\% of participants 
rated the plant-based product the same as or better 
than the animal benchmark. 
Unbreaded chicken filet (48\%) and burgers (47\%) achieved the highest shares of same or better evaluations, 
indicating near-parity for nearly half of participants. 
Breaded chicken filet (42\%), nuggets (41\%), and breakfast sausages (40\%) also demonstrated substantial competitive performance. 
Although a majority of participants 
still preferred the animal benchmark in each category, 
these results show 
that a meaningful segment of consumers 
does not perceive a clear disadvantage 
in several plant-based formats. 
\begin{figure}[h]
\centering
\includegraphics[width=1.0\linewidth]{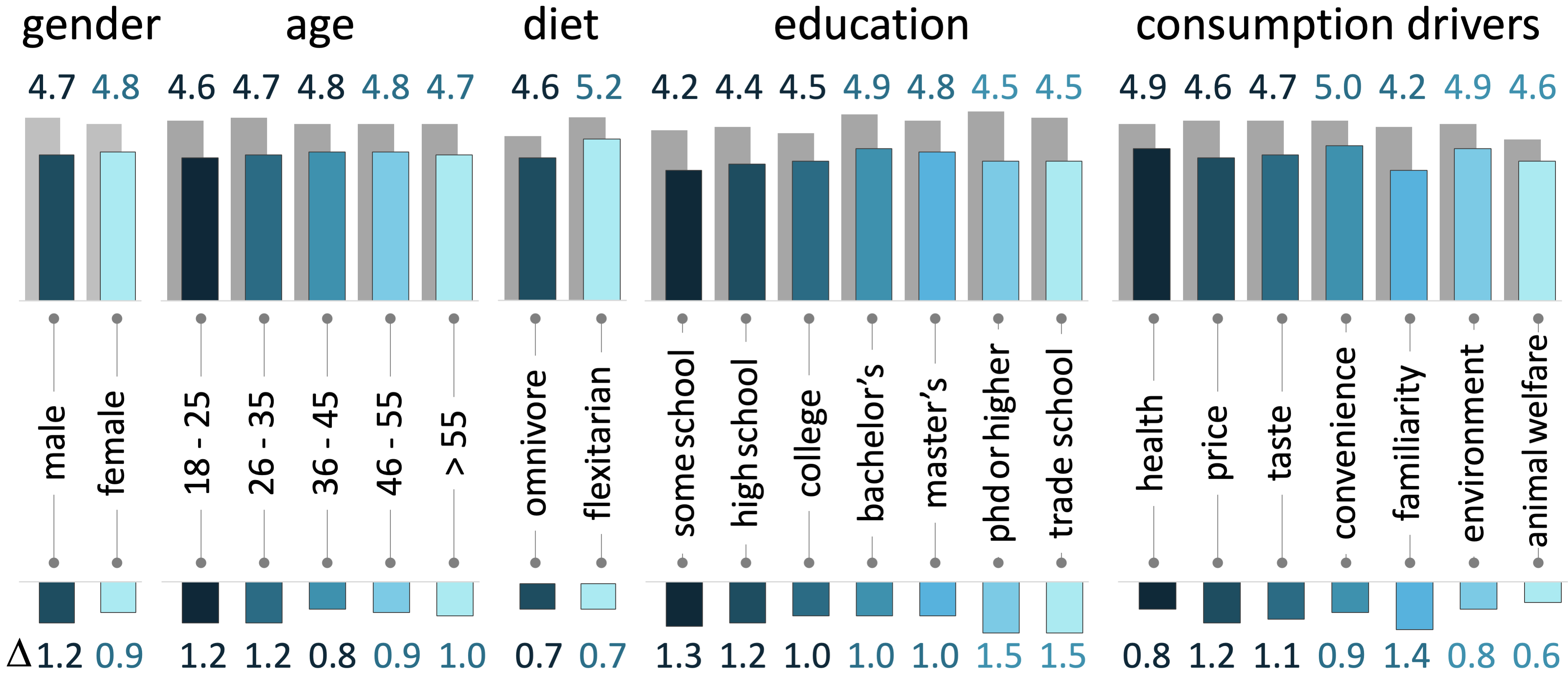}
\caption{{\bf{\sffamily{Mean purchase intent across demographic segments and consumption drivers.}}} 
Mean purchase intent for
on a seven-point Likert scale
from definitely would not buy (1) to definitely would buy (7)
across gender, age, dietary preference, education, and 
self-reported consumption drivers
for
animal benchmark (gray) and
plant-based (blue) products (top bars)
and difference $\Delta$ between both (bottom bars).}
\label{fig09}
\end{figure}
Mean purchase intent 
for animal benchmark products exceeds that 
of plant-based products 
across all demographic segments and consumption drivers 
(Fig. \ref{fig09}).
Differences between animal benchmark and plant-based products
ranged from 0.5 to 1.5 points on the seven-point scale. 
The smallest gap appears in the consumption driver of animal welfare 
(5.2 vs. 4.6; $\Delta$ = 0.5), 
whereas larger gaps appear among 
respondents with Ph.D. or trade school education 
(6.0 vs. 4.5 and 6.0 vs. 4.5; $\Delta$ = 1.5). 
Females and middle-aged consumers (36–55) showed 
relatively higher purchase intent for plant-based products 
compared with other demographic groups. 
Among consumption drivers, 
respondents prioritizing health and environment exhibited smaller differences 
($\Delta$ = 0.8), whereas those prioritizing taste, price, and familiarity showed larger preference gaps favoring animal meat ($\Delta$ = 1.1 to 1.4). 
Taken together, these results indicate 
stronger receptivity 
among health- and environment-oriented consumers, 
and middle-aged segments, and 
comparatively lower receptivity 
among younger consumers, 
and those who emphasize taste, price, and familiarity.
\begin{figure*}[h]
\centering
\includegraphics[width=0.8\linewidth]{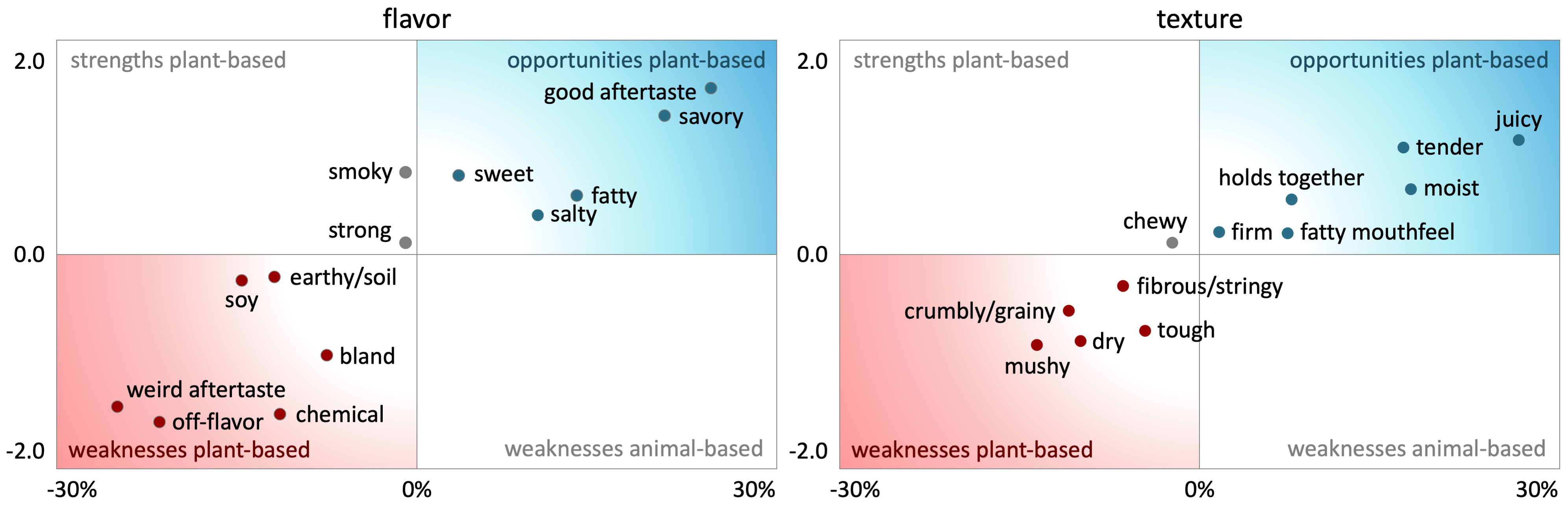}
\caption{{\bf{\sffamily{Penalty analysis of flavor and texture attributes 
and R\&D opportunities for plant-based products.}}} 
Net prevalence as percentage selecting attribute for animal minus plant-based average plotted against impact on overall liking as mean lift/penalty on a seven-point Likert scale based on check-all-that-apply responses. 
Attributes in the upper-right quadrant indicate opportunities for plant-based products;
attributes in the lower-left quadrant indicate plant-based weaknesses. 
For flavor, the largest opportunities include improving aftertaste and savoriness and reducing off-flavors and chemical notes (left).
For texture, the strongest opportunities include increasing juiciness and tenderness and reducing mushiness, dryness, and crumbliness (right).}
\label{fig10}
\end{figure*} \\[6.pt]
Penalty analysis identified 
clear sensory priorities 
for plant-based product improvement (Fig. \ref{fig10}).
We quantified each attribute 
by combining its relative prevalence 
(animal minus plant-based average) 
with its impact on overall liking. 
Attributes in the upper-right quadrant 
represent high-value opportunities 
because they occur more frequently 
in animal products and increase liking. 
In flavor, 
improving aftertaste and increasing savoriness 
produced the largest positive effects, 
while reducing off-flavors and chemical notes 
also yielded substantial gains. 
Moderate opportunities 
included increasing fattiness and saltiness 
and reducing blandness (left). 
In texture, 
increasing juiciness 
showed the strongest positive association with liking, 
followed by tenderness and moistness. 
Reducing mushiness, dryness, and crumbliness 
also improved liking 
and addressed attributes that consumers 
more often associated with plant-based products (right).
Taken together, 
these results define a prioritized sensory roadmap 
that links specific attribute gaps 
to measurable gains in overall liking.
\begin{figure}[h]
\centering
\includegraphics[width=0.8\linewidth]{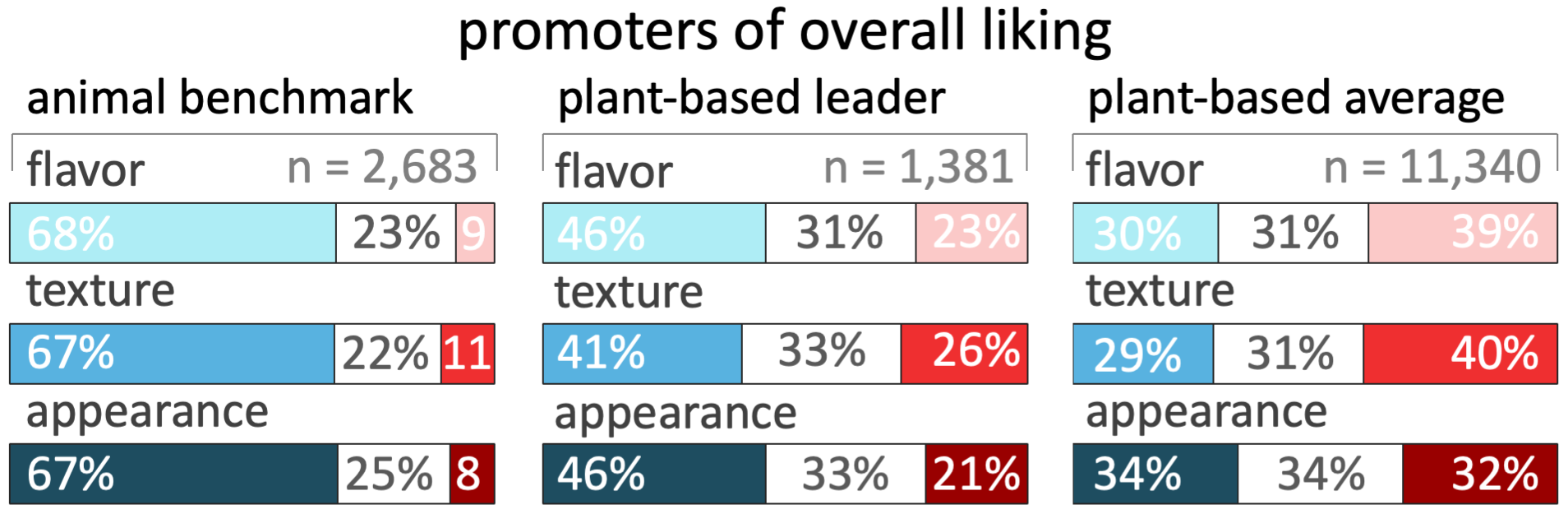}
\caption{{\bf{\sffamily{Promoter, passive, and detractor shares for flavor, texture, and appearance across product tiers.}}} 
Stacked bars show the percentage of responses 
classified as promoters (like very much and like), 
passives (like somewhat and neither like or dislike), and 
detractors (dislike somewhat, dislike, dislike very much)
for each product category 
for the animal benchmark (n = 2,683), 
plant-based leader (n = 1,381), and 
plant-based average (n = 11,340).} 
\label{fig11}
\end{figure} \\[6.pt]
Promoter, passive, and detractor shares for flavor, texture, and appearance across the animal benchmark, plant-based leader, and plant-based average 
suggest that improvements to flavor and texture should be
prioritized (Fig. \ref{fig11}).
We classified responses as
promoters (like very much and like), 
passives (like somewhat and neither like or dislike), and 
detractors (dislike somewhat, dislike, dislike very much)
based on seven-point overall liking ratings. 
For flavor, the share of promoters increases 
from 30\% for the plant-based average 
to 46\% for the plant-based leader, 
approximately a 1.5$\times$ fold increase, and 
represents the largest promoter gain across the three attributes. 
For texture, we observe the largest opportunity 
for the plant-based leader with 41\% promoters 
to close the remaining gap to the animal benchmark
with 67\% promoters.
For appearance,  
we identify the lowest R\&D priority 
because the plant-based average
receives better promoter scores for appearance with 34\%
than for flavor with 30\% and texture with 29\%.
\begin{figure}[h]
\centering
\includegraphics[width=0.8\linewidth]{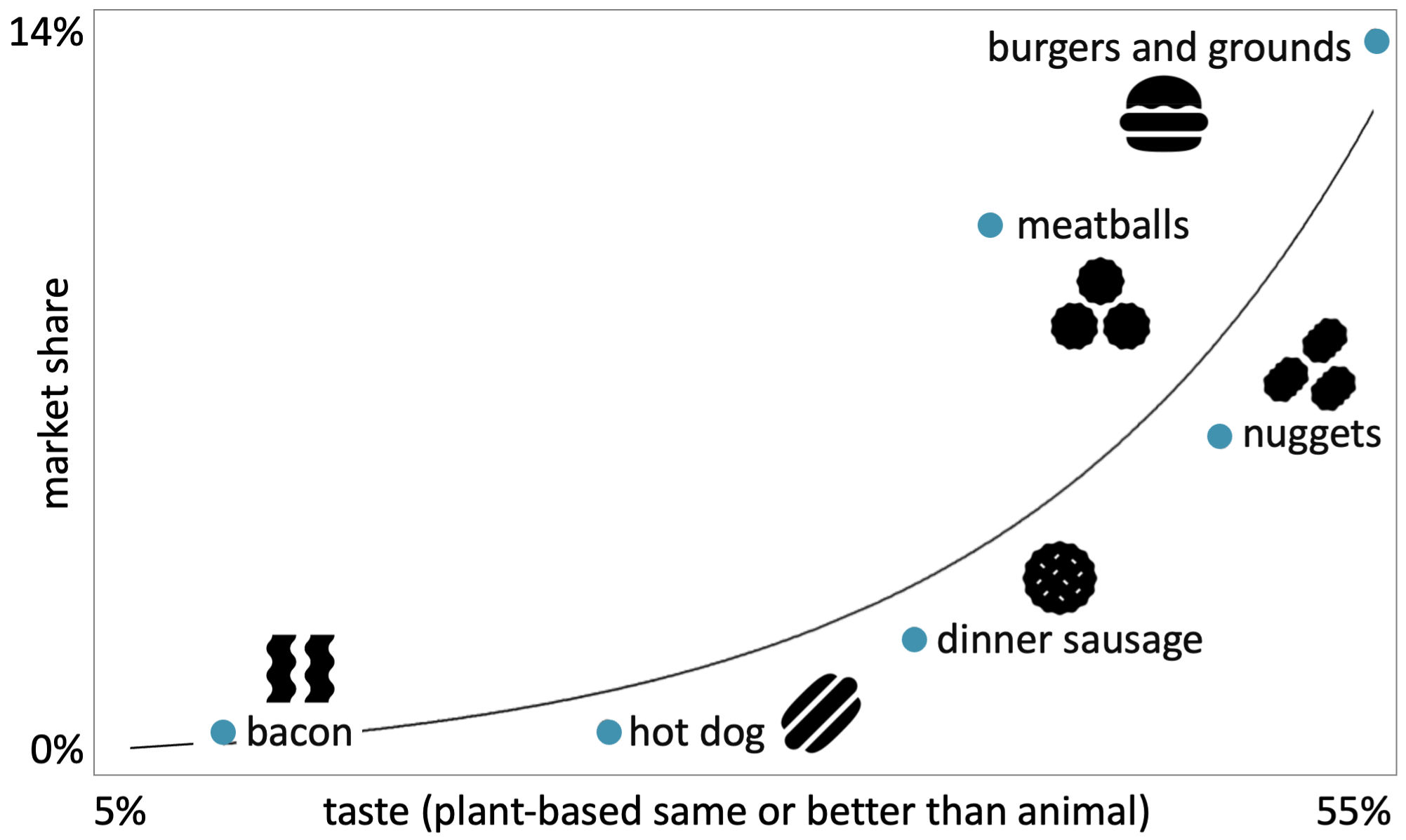}
\caption{{\bf{\sffamily{Relationship between market share and sensory performance across plant-based categories.}}} 
Scatterplot shows plant-based market share (percentage of total category sales captured by plant-based products in 2024 retail data) as a function of sensory performance (percentage of consumers rating plant-based products same or better than animal benchmark in overall liking). 
Categories with higher sensory parity exhibit substantially greater market penetration. 
Burgers and grounds, meatballs, and nuggets 
achieve both higher taste parity ($\approx$ 40–55\%) 
and higher market share ($\approx$ 6–14\%), 
whereas lower-performing categories in taste parity
such as bacon and hot dogs 
($\approx$ 10–25\%) only capture 
small fraction of category sales ($\approx$ 1\%).} 
\label{fig12}
\end{figure} \\[6.pt]
Our results reveal a strong positive association between sensory performance and plant-based market share across categories (Fig. \ref{fig12}).
We defined sensory performance as the percentage of consumers who rated plant-based products as same or better than the animal benchmark in overall liking and defined market share as plant-based sales divided by total category sales in 2024 retail data. Categories with a higher taste parity captured substantially greater market penetration. 
Burgers and grounds achieved the highest taste parity and the highest market share ($\approx$ 55\% parity; $\approx$ 14\% share), followed by meatballs ($\approx$ 40\% parity; $\approx$ 10\% share) and nuggets ($\approx$ 50\% parity; $\approx$ 6\% share). 
In contrast, bacon and hot dogs achieved substantially lower taste parity and only a marginal faction of sales 
($\approx$ 10–25\% parity; less than 1\% share). 
The fitted curve illustrates positive association between improved sensory performance and category-level penetration.
Taken together, 
these results indicate that categories that approach sensory parity secure disproportionately greater share from the animal segment, whereas categories that fail to match animal taste remain commercially constrained.
\section{{\textsf{\textbf{Discussion}}}}
This study presents 
one of the largest, blinded, in-person sensory evaluations 
of plant-based meat products to date with 
2,684 consumers evaluating products across 
14 categories resulting in 
more than 11,000 product evaluations
and more than 800,000 data points. 
Compared to our previous study 
with 1,150 consumers evaluating products across 5 categories in a test kitchen setting \cite{nectar24}, 
the current study 
more than doubled the number of participants, 
more than doubled the number of categories, 
and used authentic restaurant locations \cite{nectar25}.
At the category average level, 
plant-based products continue to lag animal benchmarks in overall liking, which aligns with prior work documenting persistent sensory gaps 
\cite{elzerman13,hoek04,hoek11}.
At the same time, 
our results show that taste parity lies within reach. 
Several plant-based products achieve strong overall liking, 
and some show no statistically significant difference from animal benchmarks in direct comparison. 
These findings support emerging evidence 
that next-generation plant-based meats 
can narrow historical acceptance gaps 
when manufacturers prioritize sensory performance
\cite{michel21,sogari23}.\\[6.pt]
{{\textsf{\textbf{Leading products highlight strong potential to meet taste expectations.}}}}
Our data show that taste parity no longer represents an abstract goal. In four cases, the animal benchmark does not achieve a statistically significant preference over a leading plant-based product. At the category level, five plant-based categories—unbreaded chicken filets, burgers, breaded chicken filets, nuggets, and breakfast sausages—achieve same or better ratings from at least 40\% of participants. Unbreaded chicken filets and burgers approach parity for nearly half of respondents. Prior research demonstrates that consumers judge plant-based meats in direct comparison to familiar animal analogues, \cite{elzerman13,hoek11}, 
and our findings indicate that several categories now compete meaningfully on taste within that comparative frame.
Especially plant-based burgers continue to draw a lot of attention \cite{merces24,tac26}, mainly because of their high market share,
and our results demonstrate that their overall liking is almost en par with their animal counterpart. \\[6.pt]
{{\textsf{\textbf{Improvement is needed, worthwhile, and attainable.}}}}
Despite the strength of the category leaders, the typical plant-based product continues to underperform. Across categories, the plant-based averages generate more detractors than promoters, while animal benchmarks show the opposite pattern. Earlier research shows that taste satisfaction drives repeat purchase and long-term substitution \cite{bryant18,bryant20,hoek11},
even in hypothetical choice experiments \cite{slade18},
and our results reinforce that conclusion. We identify clear whitespace in categories such as bacon, steak, and unbreaded chicken strips, where detractor shares remain high. At the same time, categories that achieve higher taste parity capture substantially greater market penetration. Burgers, nuggets, and meatballs reach market shares between 5\% and 14\%, whereas lower-performing categories such as bacon and hot dogs remain below 1\%. Leading products also achieve nearly double the promoter share of category averages. These results demonstrate that sensory improvement offers measurable commercial upside and remains technically achievable.\\[6.pt]
{{\textsf{\textbf{Closing the sensory gap requires decoding flavor and structure.}}}}
Flavor emerges as the dominant perceptual differentiator, 
while texture remains the most persistent structural gap 
between plant-based and animal products. 
Participants describe plant-based products 
as less savory and report off-flavors or atypical aftertastes 
more frequently than for animal benchmarks. 
Prior sensory research identifies flavor authenticity and textural realism as central determinants of meat acceptance 
\cite{elzerman13,michel21}, 
and our data reinforce this conclusion across multiple categories. 
Texture deficits--particularly reduced juiciness and altered tenderness--further diminish liking, consistent with long-standing evidence that mouthfeel and mechanical response shape perceived quality 
\cite{stpierre24,szczesniak02}. 
These findings underscore a fundamental scientific challenge: 
the need to quantitatively connect formulation, microstructure, rheology, and multisensory perception \cite{datta25}. Meat perception does not arise from a single attribute but from the integration of chemical composition, mechanical response, fat–water distribution, and flavor–aroma interactions \cite{spence15}. Bridging this gap requires models that relate measurable physical properties to sensory outcomes and that capture nonlinear, multivariate dependencies across ingredients, processing conditions, and human perception \cite{godschalk22}.\\[6.pt]
{{\textsf{\textbf{Artificial intelligence enables quantitative integration of structure and perception.}}}}
Recent advances in artificial intelligence offer a framework to address this complexity \cite{thomas25}. Multimodal learning can integrate ingredient lists, nutritional vectors, rheological measurements, and sensory descriptors into shared latent representations that enable both prediction and inverse design \cite{kuhl25}. Physics-informed neural networks and automated model discovery approaches can link deformation–stress data to perceived texture \cite{vervenne25} and translate mechanical fingerprints into interpretable structure–function relationships \cite{stpierre23}. When paired with large-scale, blinded sensory evaluations, such approaches move beyond trial-and-error reformulation toward systematic, data-driven understanding and generative artificial intelligence for systematic formulation design \cite{tac26}. Our results therefore extend beyond benchmarking current products. By curating and openly sharing more than 11,000 comparative sensory evaluations across 14 categories, we provide structured, labeled data that can support quantitative modeling of sensory performance at scale \cite{nectar25}. Adoption of sustainable proteins will depend on achieving robust sensory equivalence, as prior behavioral studies consistently demonstrate \cite{bryant18}. Increased awareness alone does not ensure dietary change \cite{thomas25a}. Instead, progress toward this goal requires mechanistic insight rather than incremental iteration. Integrating sensory science, quantitative mechanics, and artificial intelligence creates a pathway to decode what makes meat meaty and to translate that knowledge into principled advances in sustainable protein design.
\section{{\textsf{\textbf{Conclusions}}}}
This study delivers one of the largest blinded, in-person sensory comparisons of plant-based and animal meat products to date and clarifies where the field stands today: 
First, most plant-based products still require significant research and development. Across 14 categories, the plant-based average achieved 30\% promoters compared with 68\% for animal benchmarks and generated nearly four times as many detractors. 
Second, leader products have emerged in most categories. In unbreaded chicken filet, burgers, nuggets, breaded chicken filet, and breakfast sausage, at least 40\% of participants rated the plant-based leader the same as or better than the animal benchmark. 
Third, taste parity is achievable. In unbreaded chicken filet, the within-subject comparison revealed no statistically significant preference for the animal product (p = 0.314), and the overall liking gap reached only 0.1\,-\,0.2 points in the best-performing categories. 
Fourth, sensory performance strongly correlates with commercial performance: categories that approach taste parity, such as burgers, nuggets, and meatballs, secure 5\,-\,14\% market share, whereas low-parity categories such as bacon and hot dogs remain below 1\%. 
Finally, the path to parity requires targeted improvements in flavor and texture. Penalty analysis identifies savoriness, aftertaste, juiciness, and tenderness as the most powerful levers to increase liking, while appearance plays a secondary role relative to flavor and texture.\\[6.pt]
Taken together, these findings show that plant-based meat has moved beyond proof of concept. The category now contains clear sensory leaders and commercially viable formats; yet, it has not achieved consistent mainstream equivalence across categories. Plant-based meat stands at a transitional moment: Technical feasibility exists, category-level performance varies widely, and systematic sensory optimization can close the remaining gap. To accelerate progress, we make all sensory, preference, and market-linked data from this study publicly available. By democratizing access to large-scale, blinded sensory evidence, we aim to enable independent validation, stimulate rigorous food science, and accelerate discovery and innovation across the sustainable protein ecosystem.
\section*{CRediT authorship contribution statement}
{\bf{Sybren D. van den Bedem:}} 
Writing -- original draft, 
Writing -- review \& editing, 
Visualization, 
Formal analysis.
{\bf{Ellen Kuhl:}} 
Writing -- original draft, 
Writing -- review \& editing,
Visualization, 
Formal analysis.
{\bf{Caroline Cotto:}}
Writing -- original draft,
Writing -- review \& editing,
Visualization,
Validation, 
Supervision, 
Project administration, 
Methodology, 
Investigation,
Formal analysis, 
Data curation,
Conceptualization. 
\section*{Ethical statement}
This study was conducted in collaboration with Palate Insights, 
a San Francisco-based company specializing in food surveys,  
in accordance with Stanford University 
Institutional Review Board guidelines. 
\section*{Funding statement}
This research was supported 
by 
Food System Innovations (Humane America Animal Foundation) and
by seed funding 
from the Stanford Bio-X Snack Grant and
from the Stanford Doerr School of Sustainability Accelerator, and
by the NSF CMMI grant 2320933. 
\section*{Declaration of competing interest}
The authors declare that they have no known competing financial
interests or personal relationships that could have appeared to influence
the work reported in this paper.
\section*{Data availability}
All data are freely available at 
https://www.nectar.org/ sensory-research/2025-taste-of-the-industry.
\section*{Acknowledgements}
The authors would like to thank
Alex Weissman, Han Gu, and Tom Conger 
from Palate Insights, 
and Max Elder from Food System Innovations.

\end{document}